External noise removed from magnetoencephalographic signal using Independent Component Analyses of reference channels

Jeff Hanna[a] , Cora Kim[a] , Nadia Müller-Voggel[a]
[a]University Clinic – Erlangen, Neurosurgery
Erlangen, Germany

Abstract

Background:
Many magnetoencephalographs (MEG) contain, in addition to data channels, a set of reference channels positioned relatively far from the head that provide information on magnetic fields not originating from the brain. This information is used to subtract sources of non-neural origin, with either geometrical or least mean squares (LMS) methods. LMS methods in particular tend to be biased toward more constant noise sources and are often unable to remove intermittent noise.

New Method:
To better identify and eliminate external magnetic noise, we propose performing ICA directly on the MEG reference channels. This in most cases produces several components which are clear summaries of external noise sources with distinct spatio-temporal patterns. We present two algorithms for identifying and removing such noise components from the data which can in many cases significantly improve data quality.

Results:
We performed simulations using forward models that contained both brain sources and external noise sources. First, traditional LMS-based methods were applied. While this removed a large amount of noise, a significant portion still remained. In many cases, this portion could be removed using the proposed technique, with little to no false positives.

Comparison with existing method(s):
The proposed method removes significant amounts of noise to which existing LMS-based methods tend to be insensitive.

Conclusions:
The proposed method complements and extends traditional reference based noise correction with little extra computational cost and low chances of false positives. Any MEG system with reference channels could profit from its use, particularly in labs with intermittent noise sources.

Introduction

Neurophysiological data contain many sources of noise that are orders of magnitude stronger than the signal, requiring substantial cleaning before they can be properly analysed. A non-exhaustive list of methods includes: hand removal of bad sections of data, temporal filtering, regression, blind source separation, signal space projection, beamforming, canonical correlation analysis, wavelet transforms, and empirical-mode decomposition (for recent reviews, see Urigüen and Zapirain, 2015, and Mannan et al, 2018). The method described here belongs in the blind source separation (BSS) family. Briefly, BSS methods 1) factor neurophysiological data into distinct, underlying components, then 2) the artefactual components are identified and removed, and 3) the remaining components are reassembled back into data-space. Within step (2) itself, there are a variety of approaches for how best to identify bad components. We propose here a method that is based on decomposition of the reference channels into distinct sources, and using those as a form of ground truth for determining which sources should be removed from a full ICA decomposition that includes both reference and standard channels.

Prior use of reference channels in magnetoencephalography

Even with the use of magnetically shielded rooms, there remains an enormous amount of environmental magnetic noise in the magnetoencephalographic (MEG) signal. One of the most common means of removing this noise is by positioning an array of sensors further from the head than the standard sensors. Because the strength of a magnetic field follows an inverse power law of the distance, and because the magnetic fields originating from the brain are many orders of magnitude weaker than those from the environment, there is a high degree of certainty that whatever is measured in the reference channels is not of neural origin. However, this degree of certainty is conditioned on the obligatory compromise made with the distance of the reference channels from the standard channels: the closer the two types of channel arrays are, the better noise estimation the reference channels provide of the spatially inhomogeneous magnetic noise fields, but also the more chance that brain signals are picked up in the reference channels.

The reference channels then provide information on the ambient magnetic noise in the environment that can be used to remove that noise from the data channels. Typically, a set of weights are derived which map the relationship between a signal in the reference channels to that same signal's expression in the standard channels. These weights are then used to subtract that signal from the standard channels. This can be done geometrically (e.g. Vrba, 1996) or by using least mean squares (LMS) (Widrow et al, 1975). While LMS approaches significantly reduce general, background noise, they do not perform very well with intermittent patterns, (i.e. machines starting and stopping), owing to the fact that the weights are calculated from the entire recording, and so naturally orient themselves toward the more constant noise sources. To remedy this, refinements to LMS approaches have been proposed where the weights are updated continuously in user-specified intervals throughout the length of the recording (Adachi et al, 2001, Ahmar and Simon, 2005).

Proposed method

The method we introduce here is very similar in its logic to the well-established practice of removing eye- or heart-related artefacts by performing independent components analysis (ICA) on a magneto/electroencephalographic (M/EEG) signal and removing those components which have a spatio-temporal pattern consistent with ocular/cardiac activity (Joyce et al, 2004). ICA is a powerful form of blind source separation, which can, with minimal assumptions and without any prior information, separate a signal with N concurrent channels into as many as N underlying components (Hyvärinen and Oja, 2000). More precisely, for every n of N components, C channels, and T time points, a weight vector w of length C is found that, when multiplied against the CxT data matrix, results in a 1xT component time course that 1) is completely uncorrelated with any other component time course, and 2) has a distribution which is minimally Gaussian. With only these two constraints, distinct spatio-temporal patterns in the data can be separated from each other and manipulated at will without affecting the other patterns. When applied specifically to M/EEG data, ocular-related noise is usually cleanly encapsulated into one or two independent components (ICs), and in MEG, cardiological noise is also often well-captured. The ICs for both of these noise sources have such typical spatio-temporal properties that an experienced analyst can identify them by visual inspection without much trouble, though it may be desirable instead to correlate the components against a ground truth signal, such as an electro-oculagram (EOG) or -cardiogram (ECG), which gives independent, reliable information about the activity of the eyes/heart. This approach has the advantage of removing any potential for subjective bias on the part of the analyst, while also sparing them some tedious labour.

The innovation here is the use of a new source of ground truth analogous to EOG/ECG channels, but oriented to external magnetic noise instead, derived by performing an ICA on the reference channels only. This ICA decomposition isolates noise patterns with distinct spatio-temporal profiles, where the spatial information is described by a component's weight distribution on the reference channels. Importantly, there is no bias in the ICA decomposition against intermittent sources; such noise bursts are in most cases cleanly isolated into their own component.

In the next step, a second ICA decomposition is performed on both the reference channels and the standard channels together. Components in the full decomposition which correlate beyond a given threshold with any of the components from the reference-only decomposition are removed, thus reducing or eliminating that noise source's contribution to the signal in the standard channels. We also explore here an alternative algorithm where only a single ICA decomposition is performed, on both reference and standard channels together, and those components whose weights are particularly heavy on the reference channels are excluded. Henceforth we refer to the former algorithm as the "separate" algorithm, and the latter as "together." They are described in full on table 1.

INSERT TABLE 1 HERE

The efficacy of the algorithms was assessed by applying them on simulated MEG data that contain both brain and external noise sources. In this way, the noise and brain contributions to the simulated signal are perfectly known, and provide an objective means of assessment, as well as some guidance for parameter selection. In addition to this, we show results on real MEG data collected in our lab, which demonstrate that the algorithms remove well-known sources of external noise.

Methods

Simulations

All simulation data consisted of two superimposed parts: a brain part, i.e. cortical sources, and a noise part, i.e. external sources. These were built in the following steps: (1) In order to keep the temporal variation of the time courses of both the brain and noise parts as near as possible to those seen in real data, we derived them from 20 real MEG recordings, comprised of four blocks for each of five participants, lasting about five minutes each. During these recordings, subjects were completing various attention-based experimental tasks. For producing the brain part time courses, the MEG data were cleaned carefully by visual inspection and then inverted with Minimum Norm Estimation (MNE) onto a distributed surface source space of 1026 sources per hemisphere, derived from an individual structural MRI, segmented with Freesurfer (Dale et al, 1999, Fischer et al, 1999,2001,2002). The noise time courses were derived by performing ICA decomposition on the same 20 recordings (Picard algorithm, Ablin et al, 2018) on only the 23 reference channels, with the constraint that only the components which explained the first 99% of the variance were calculated. This resulted in 2-6 noise components per recording, for a total of 95 noise time courses. The next step (2) is to build the source spaces for brain and noise parts. For the brain part, we already have the source space we used in step (1). For the noise part, we placed 2-6 magnetic dipoles in three dimensional space. The coordinates had a random uniform distribution between 1 and 500 meters for each of the three axes. Dipole orientations were also randomly set with a uniform distribution.

In step (3), we assign the time courses derived in step (1) to the source spaces from step (2). For the brain part, again the source time course is assigned to the same source space dipole which produced it in in step (1). In the noise part, for each of the 2-6 magnetic dipoles, we randomly assigned it one of the 95 possible noise time courses derived in step (1). The noise time courses - originally ICA components with an arbitrary -1 to 1 range - were scaled into SI unit Teslas with the following formula.

$$y = xAr^3 10^{-7}$$

where x is the original time course, r is the distance of the dipole from the origin, A is a random variable with a normal distribution of mean 1 and standard deviation of 0.1, and $10^{-7}$ is a constant that was arrived at by trying different values until the reference channel magnitudes closely matched those typically seen in real recordings. The cubed distance multiplier ensures that distant sources continue to have a measurable effect on the simulated data, and the random variable effects small, random increases/decreases in the signals. One further adjustment was made to the scaling with the purpose of simulating the effects of the magnetically shielded room (MSR). Sources outside the simulated MSR - approximated by a sphere with a radius of 3m - were first Fourier transformed, and the resulting Fourier space signal was convolved on both the real and imaginary axes with a log linear function, beginning with a suppression factor of 100 at 0Hz and ending with a factor of 100,000 at 100Hz. This closely approximates the suppression/Hz function of our MSR (Vakuumschmelze, Hanau). However, after this filtering, the signals would be too weak to be measured, so they are multiplied again with a factor that equalises the sum of squares of the filtered signal with the sum of squares of the pre-filtered signal. In this way, the amplitude is kept at a level which can be detected in the sensors, while ensuring the spectral properties induced by the MSR are preserved in simulated data. Sources inside the MSR i.e. with a distance of less than 3 meters from the sensor origin were left unfiltered.

For each distinct noise part we produced, there was a minimum of 2 and maximum of 4 noise sources outside the MSR, and a minimum of 0 and maximum of 2 sources inside the MSR. We produced in total 100 distinct noise parts, each with an accompanying forward model that describes the relationship between the channels and the noise sources. Figure 1 shows a schematic (not to scale) depiction of the source model for a brain part along with the source model for a noise part, as well as both data channels and the reference channels (excluding the gradiometers).

In the final step (4), we use the source spaces, time-courses, and forward models to produce simulations of the 20 possible brain parts, and the 100 possible noise parts. Because of the superposition principle, brain and noise parts can be simply added together to achieve the effect of data with both cortical and external sources. We added together all possible combinations of brain and noise parts, for a total of 2000 simulated data sets. A first-pass noise reduction was applied to the simulated data using a standard LMS approach (see Introduction). In particular, a set of coefficients was calculated for the reference channels that best predict the signal in the standard channels. This predicted signal was then subtracted from the standard channels, leaving only those parts not predictable by the reference channels - i.e. brain activity and also parts of the external noise not captured by LMS. To ensure that absolutely no residual brain signal in the reference channels could influence the LMS calculation, it was estimated only on the noise part, i.e. before it was added together with the brain part. The simulations were created using a version of MNE-Python v.20 (Gramfort et al, 2013,2014) that was modified slightly so that forward models/simulations can include reference channels.

INSERT FIGURE 1 HERE

Real data

The MEG data shown here were taken from an unrelated experiment, where participants were recorded in supine position while they were presented with pictures of varying emotional valence. The MEG system used was a 4D Magnes 3600, with 248 magnetometers and 23 reference channels. Data were high-pass filtered online at 1Hz and recorded with a sampling rate of 678.17 Hz. Reference channel-based noise cancellation was applied online; the coefficients used in our MEG system are calculated by a CTF (Coquitlam, BC, Canada) technician – this is done biannually to compensate for slow changes in the ambient, stationary magnetic noise fields. The data were notch filtered offline for line-noise (50Hz and harmonics) as well as at 62Hz for a persistent noise source of unknown origin at this frequency. Data were then downsampled to 200Hz. Any rare cases of large electromyographic (EMG) noise or technical disturbances were identified and removed by visual inspection. ICA and the proposed algorithms were then applied and EOG and ECG components were selected by visual inspection before finally applying the proposed technique.

Assessment

Significance of all correlations was determined with the iterated Z-scoring technique in MNE Python (Gramfort et al, 2013, 2014): 1) for a given noise source, the Pearson r scores are calculated against each component, 2) these r scores are transformed to Z scores. 3) Z scores above a threshold are marked as bad and removed 4) Z scores are recalculated. Steps 3 and 4 are repeated until no Z scores are above threshold. Correlation thresholds are therefore specified henceforth in Z scores, rather than r scores. This technique focusses rather on those correlations which are particularly high in relation to the other correlations, rather than the absolute value of the correlations.

Component identifications were divided into "hits," "misses," and "false alarms." When a component identified by the algorithm also correlated with the noise source beyond a threshold, this was tallied as a "hit." A component which correlated with a noise source beyond threshold but was not identified by the algorithm was tallied as a "miss." A component identified by the algorithm that did not also correlate beyond threshold with a noise source was tallied as a "false alarm." We tallied hits/misses/false alarms for both algorithms with a range of different Z score thresholds. In all cases however, the correlation threshold of components directly against noise sources was held constant at Z=3.

Rank estimation and number of independent components

We also assessed the performance of the algorithms as a function of the number of components used in the ICA decomposition. In order to determine which range of component numbers should be tested, we performed rank estimation on a random sample of 200 of the 2000 simulated data sets. This was done with two methods. The first was the Scree plot/Kaiser criterion method, where the eigenvalues of the principal component analysis (PCA) are calculated, and all components with eigenvalues of less than one are considered superfluous. The second was the average log-likelihood of the data under the probabilistic PCA framework (Tipping and Bishop, 1999), calculated within a 5-fold cross validation. With the first method, the eigenvalues tended to go under the unit threshold at the 23rd component. With the latter method, the log-likelihood increased monotonically throughout the whole range of tested component numbers, but began to asymptote also shortly after 20 components. On the basis of these results, we tested the algorithms with 20,40,60,80, and 100 components. Dimension estimation was performed for both the data channels alone, and for the data and reference channels together, with the predictable result that reference channels tended to add a few dimensions of rank to the data. Estimations are depicted in fig. 2.

Results

Simulations

Component identification performance is depicted for both algorithms in figure 3. A few properties become immediately clear. With the "separate" algorithm, both sensitivity and accuracy are optimal at 20 components - which for these data was very close to their estimated rank - and degrades moderately with the addition of more components. The optimal Z threshold would be 3 or 3.5, depending on whether sensitivity or accuracy is more important to the researcher, respectively. With the "together" algorithm, sensitivity and accuracy are also generally better with less components than more components, though performance at higher Z thresholds is quite poor with 40 and especially with 20 components. A threshold of Z=2.5 for the "together" algorithm seems optimal, whatever the number of components.

INSERT FIGURE 2 HERE

Real Data Example

There are several known, intermittent noise sources that have been consistently observed at our MEG facility at the University Clinic - Erlangen. Here, we show an example of a particularly unlucky recording that contained all of them, and how well the proposed algorithm performs under these conditions. These sources include 1) a powerful, transient burst at about 1.5Hz and its harmonics, 2) train noise at 16.6Hz, and 3) a dual-peak 23,24Hz noise that comes and goes throughout the recording. Noise sources 1 and 3 are of unknown origin, though they have been regularly observed in our lab for many years and are assumed to originate from machines in or around the facility. The rank of the data was estimated at around 60, and so ICA was performed with as many components. We then applied the "separate" algorithm with Z=3, and were able to remove the majority of these noise sources from the data.

Figure 4 shows the power spectrum densities (PSDs) for the data before the algorithms were applied (but after all other cleaning procedures were applied, including online noise correction, EOG- and ECG-based cleaning, see Methods), and after the "separate" and "together" algorithms were applied. Noise source 1 seems to have been entirely removed, and 3 is significantly attenuated. Noise source 2 (the train) is eliminated in a subset of the channels, but remains as strong in the others. Figure 5a shows the topographies of the ICA components selected by the algorithm, 5b shows their time courses, and 5c shows their spectral properties. The temporal and spatial properties of these components illustrate what kinds of noise the technique can catch – and what kinds are missed by traditional methods. Components 0 and 35 for example reflect a noise burst which occurs only in the first ~1/4 of the recording and is afterwards completely silent. Component 4 reflects cardiac activity that we did not initially identify when cleaning by hand. The PSDs make clear that component 20 is the 23,24Hz noise, and the segments image of that component shows that it comes and goes in regular intervals throughout the recording. Component 57 and to a lesser extent 35 seem to reflect the 16.6Hz train noise. It is not immediately clear what noise sources are captured in components 39 and 41, though their weight topographies are broadly consistent with the topographies of external noise we are accustomed to seeing in our laboratory.

INSERT FIGURES 4 AND 5 HERE

Discussion

We have proposed a novel technique that identifies and removes external noise through an ICA decomposition of the reference channels, and developed two algorithms to realise this. Application of the technique to simulations of MEG data which included external noise alongside brain data have shown convincingly that both algorithms remove a large amount of external noise which is not removed by traditional LMS noise subtraction, and do so with little to no false positives. The "together" algorithm is the more conservative of the two, in the sense that it is very unlikely to produce false positives, but also less sensitive to actual sources of noise. We would nevertheless caution against a too narrow interpretation of the simulation results, or the adoption of sensitivity parameters based on them alone. The proposed technique is ICA-based, and exactly how ICA partitions the signal into components can vary in ways that are not always straightforward to predict. This in turn makes it difficult to build simulations that fully encompass the variability one may see in different electro-magnetic environments, different equipment, different participant populations, etc. Also, our modelling of the MSR in these simulations was only approximate, and we cannot therefore rule out that there are finer-grained MSR distortions in real data that were not present in the simulations. A further qualifier is that the tallying of hits/misses/false alarms depends necessarily on the threshold set for how much the components may correlate directly against the noise in order to be considered noise-related. The $Z=3$ threshold we used seems to be a good compromise on sensitivity/accuracy, but whatever value we chose would have to have been to some extent arbitrary. We recommend therefore that researchers use the simulation results as a form of rough guidance for parameter selection, rather than as a final determination. In practice, for example, we have found in our lab that using $Z=3$ with the "separate" algorithm on real data captures only components that are obviously of non-neural origin, even though the simulation data would suggest that at these settings there are a small but non-trivial amount of false alarms.

Aside from parameter selection, the simulation results provide guidance on the overall tendencies of the algorithms. Provided enough components are used (i.e. >40), the "together" algorithm has consistent performance across a wide range of parameters, while the "separate" algorithm merits more careful parameter selection. Researchers should weigh these properties against their individual circumstances and goals when deciding which one to adopt. We would also highly recommend limiting the number of ICA components to the estimated rank of the data.

One particularly salient issue here is the question of moving noise sources. The technique proposed here seems to function adequately with spatially fixed noise sources. Our simulations did not include moving sources, so we cannot comment directly on how it would perform with them, though there are reasons to suspect it may be sub-optimal. An ICA component is a vector of weights that, in the context of MEG, constitute a distinct and fixed spatial pattern. Moving sources will however continually change their distribution on the channels, and cannot be captured by any one set of weights. Indeed, as shown in the real data example, the technique performs worst with the train, which is the one noise source that we know to be moving. MEG laboratories which are particularly burdened by large, moving noise sources may then rather opt for continually adjusted weight methods mentioned in the introduction (Adachi et al, 2001, Ahmar and Simon, 2005). A good avenue for further development would be to adapt this technique to use multiple ICA decompositions throughout the recording in user-specified windows.

Another potential area of future development is in the use of the proposed method for removal of artefacts associated with deep brain stimulation or vagus nerve stimulation. We did not simulate these here, owing to the significant added complexity of modelling the stimulator dipoles. Nevertheless, evaluating the sensitivity of the proposed method to these types of artefacts - whether with real data or simulations - would be a worthwhile project with potentially significant clinical application.

Conclusion

We have proposed a conceptually simple technique for removing intermittent external noise from magnetoencephalography, using ICA decomposition on reference channels. In many cases, this will provide significant improvements to the data quality. Two algorithms are available which have different sensitivity/accuracy trade-offs. Their efficacy is demonstrated on simulated data, which also provide qualified guidance on algorithm and parameter selection.

Acknowledgements

We would like to thank Antonia Keck, Denise Kunze, and Carolin Spielau-Romer for assistance in data collection, and Martin Kaltenhäuser, Stefan Rampp, Eric Larson, and two anonymous reviewers for helpful discussion. Computing power was provided by the High Performance Computing facilities at the Friedrich-Alexander Universität Erlangen-Nürnberg. This research was funded by the Deutsche Forschungsgemeinschaft, through an Emmy Noether Grant awarded to Nadia Müller-Voggel (grant number: 334628700). Conflict of interest statement: none.

Implementation

The "separate" algorithm is currently implemented in MNE Python 0.18 or higher. The "together" algorithm is planned for later versions.

References

Ablin, P., Cardoso, J. F., & Gramfort, A. (2018). Faster independent component analysis by preconditioning with Hessian approximations. IEEE Transactions on Signal Processing, 66(15), 4040-4049.

Adachi, Y., Shimogawara, M., Higuchi, M., Haruta, Y., & Ochiai, M. (2001). Reduction of non-periodic environmental magnetic noise in MEG measurement by continuously adjusted least squares method. IEEE Transactions on Applied Superconductivity, 11(1), 669-672.

Ahmar, N. E., & Simon, J. Z. (2005, March). MEG adaptive noise suppression using fast LMS. In Conference Proceedings. 2nd International IEEE EMBS Conference on Neural Engineering, 2005. (pp. 29-32). IEEE.

Dale, A. M., Fischl, B., & Sereno, M. I. (1999). Cortical surface-based analysis: I. Segmentation and surface reconstruction. Neuroimage, 9(2), 179-194.

Fischl, B., Sereno, M. I., & Dale, A. M. (1999). Cortical surface-based analysis: II: inflation, flattening, and a surface-based coordinate system. Neuroimage, 9(2), 195-207.

Fischl, B., Liu, A., and Dale, A. M. (2001). Automated manifold surgery: constructing geometrically accurate and topologically correct models of the human cerebral cortex. IEEE Trans. Med. Imaging 20, 70–80. doi:10.1109/42.906426

Gramfort, A., Luessi, M., Larson, E., Engemann, D. A., Strohmeier, D., Brodbeck, C., ... & Hämäläinen, M. S. (2014). MNE software for processing MEG and EEG data. Neuroimage, 86, 446-460.

Gramfort, A., Luessi, M., Larson, E., Engemann, D. A., Strohmeier, D., Brodbeck, C., ... & Hämäläinen, M. (2013). MEG and EEG data analysis with MNE-Python. Frontiers in neuroscience, 7, 267.

Hyvärinen, A., & Oja, E. (2000). Independent component analysis: algorithms and applications. Neural networks, 13(4-5), 411-430.

Joyce, C. A., Gorodnitsky, I. F., & Kutas, M. (2004). Automatic removal of eye movement and blink artifacts from EEG data using blind component separation. Psychophysiology, 41(2), 313-325.

Mannan, M. M. N., Kamran, M. A., & Jeong, M. Y. (2018). Identification and removal of physiological artifacts from electroencephalogram signals: A review. *IEEE Access*, *6*, 30630-30652.

Ramachandran, P. and Varoquaux, G., (2011). Mayavi: 3D Visualization of Scientific Data. IEEE Computing in Science & Engineering, 13 (2), pp. 40-51

Urigüen, J. A., & Garcia-Zapirain, B. (2015). EEG artifact removal—state-of-the-art and guidelines. Journal of neural engineering, 12(3), 031001.

Tipping, M. E., & Bishop, C. M. (1999). Mixtures of probabilistic principal component analyzers. Neural computation, 11(2), 443-482.

Vrba, J. (1996). SQUID Sensors: Fundamentals, Fabrication and Applications. Kluwer Academic, Dordrecht, 117-123.

Widrow, B., Glover, J. R., McCool, J. M., Kaunitz, J., Williams, C. S., Hearn, R. H., ... & Goodlin, R. C. (1975). Adaptive noise cancelling: Principles and applications. Proceedings of the IEEE, 63(12), 1692-1716.

Table 1

"Separate" algorithm:
1) Perform ICA on only the reference channels
2) Perform ICA on reference and data channels together
3) Remove ICs from (2) that correlate above threshold with ICs from (1)

"Together" algorithm:
1) Perform ICA on reference and data channels together
2) Divide the L2-norm of the reference channel ICA weights by the L2-norm of the standard channel ICA weights
3) Log-transform the ratios from (2) to give them a normal distribution
4) Transform the logs from (3) into Z-scores
5) Remove components whose Z-score is above threshold

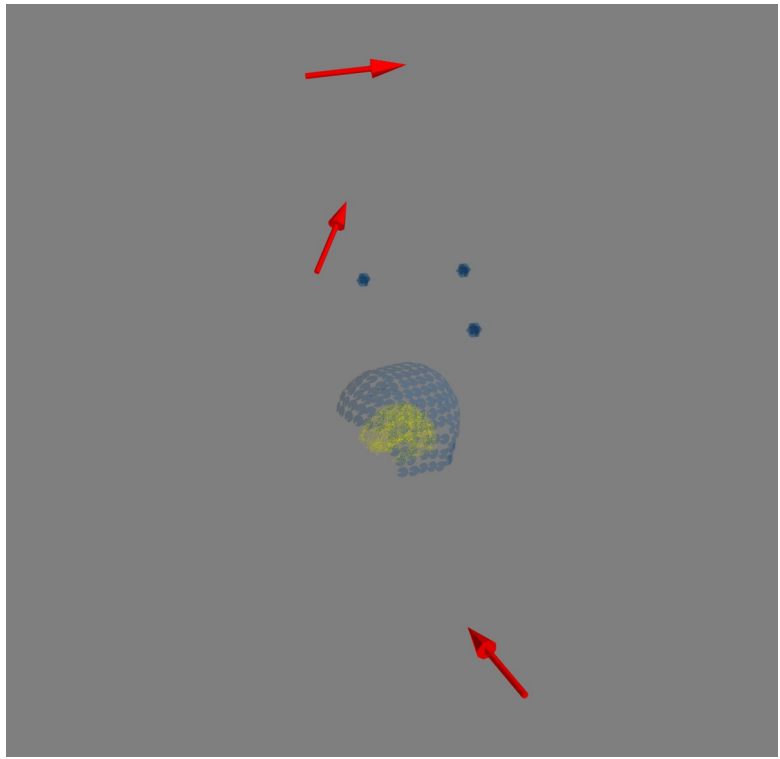

Figure 1: A schematic representation (i.e. not to scale) of one of the 2000 source models used for simulations. Here, there are 2052 brain dipoles (in yellow), and three noise dipoles (red arrows). Magnetic fields from the dipoles are projected onto the 248 magnetometers arrayed within the helmet, as well as 18 of the 23 reference channels above the helmet (the 5 gradiometers are excluded), in order to produce approximations of MEG data containing a mix of brain and noise components. Rendered with Mayavi 4.7.1 (Ramachandran and Varoquaux, 2011)

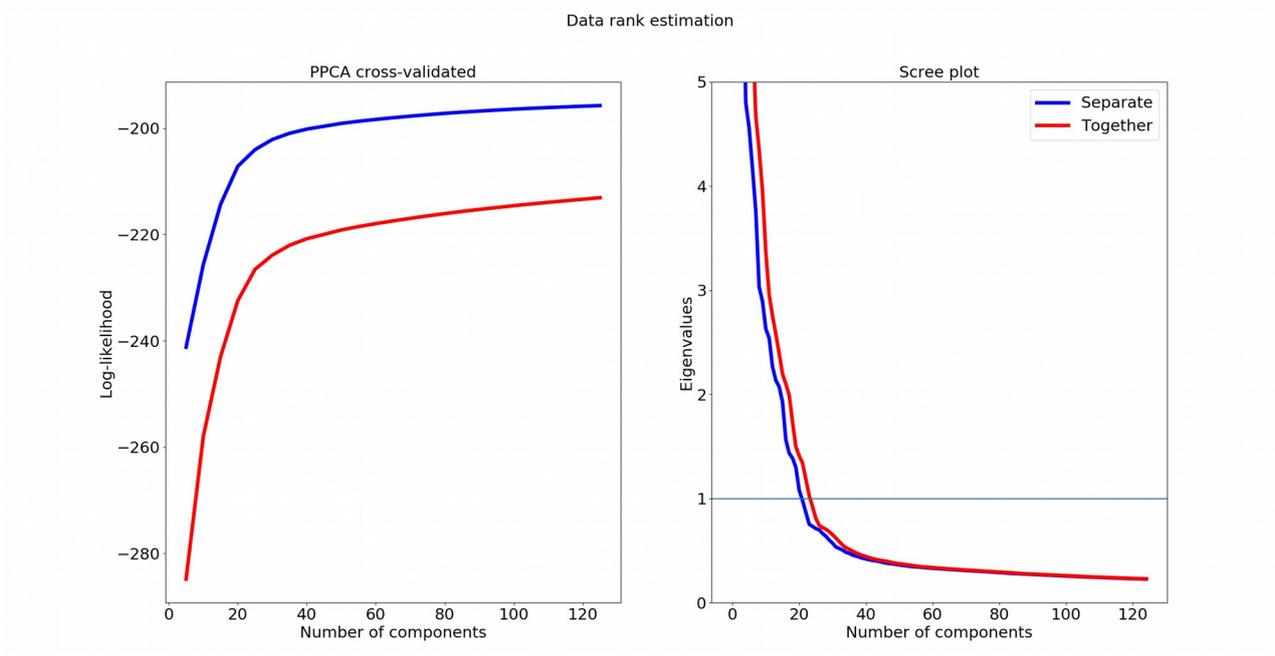

Figure 2: Estimation of the rank of simulated data using cross-validated probabilistic principal component analysis (PPCA) and a scree plot. In the scree plot, the Kaiser criterion is marked with a horizontal line. Results from both methods suggest a rank of somewhat more than 20.

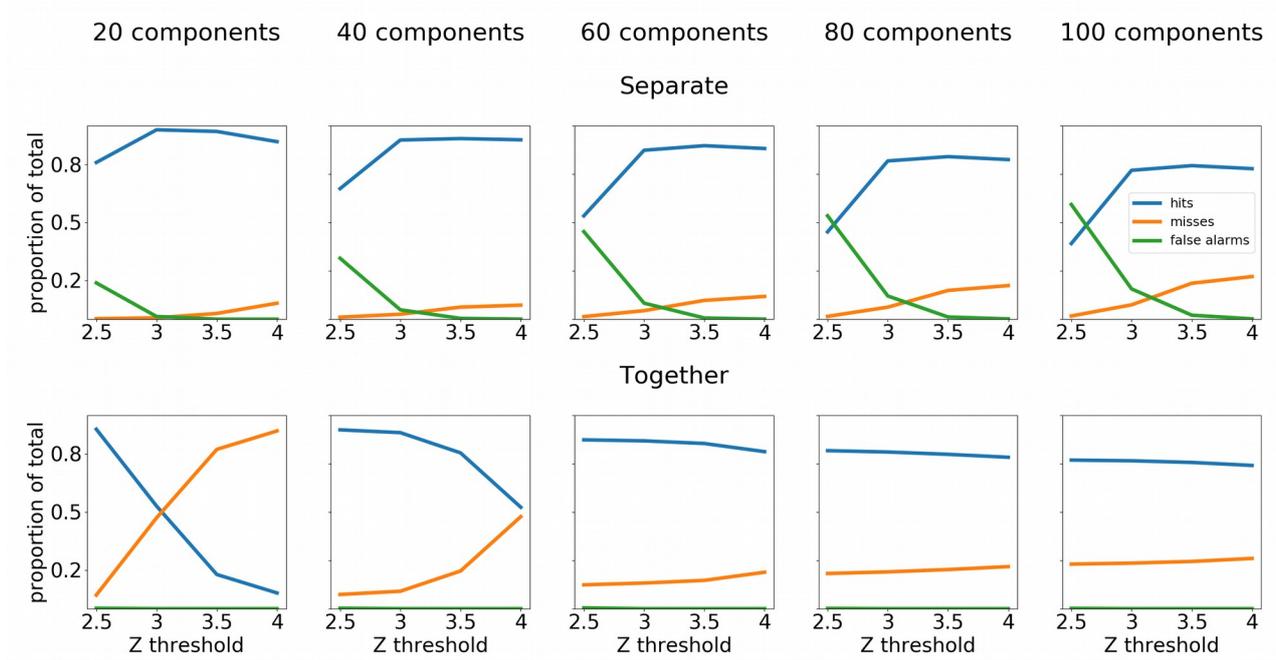

Figure 3: Performance of the proposed technique on simulated data for "separate" and "together" algorithms under different sensitivity settings (Z threshold), and with different numbers of independent components. Lines indicate the proportion of components caught, missed, or falsely identified by the technique.

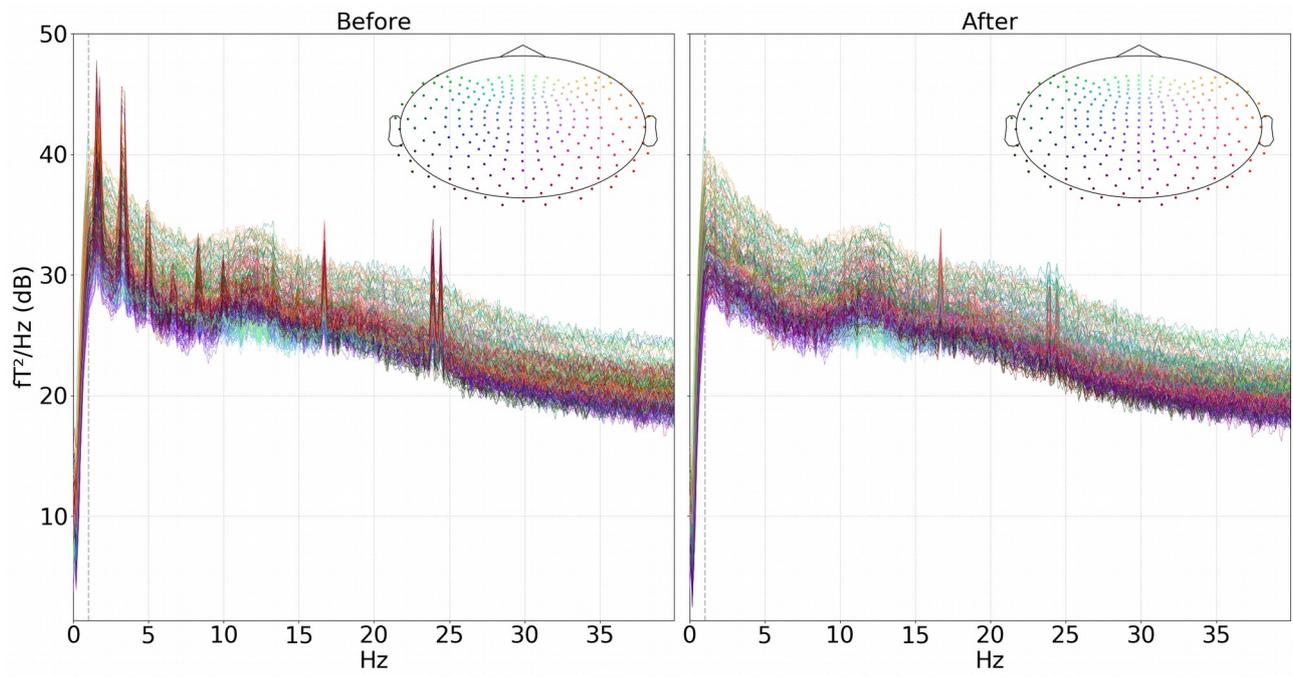

Figure 4: Power spectrum densities (PSD) for a real MEG recording before and after the "separate" algorithm was applied with Z=3. The many, sharp peaks in the "before" picture reflect well-known noise sources.

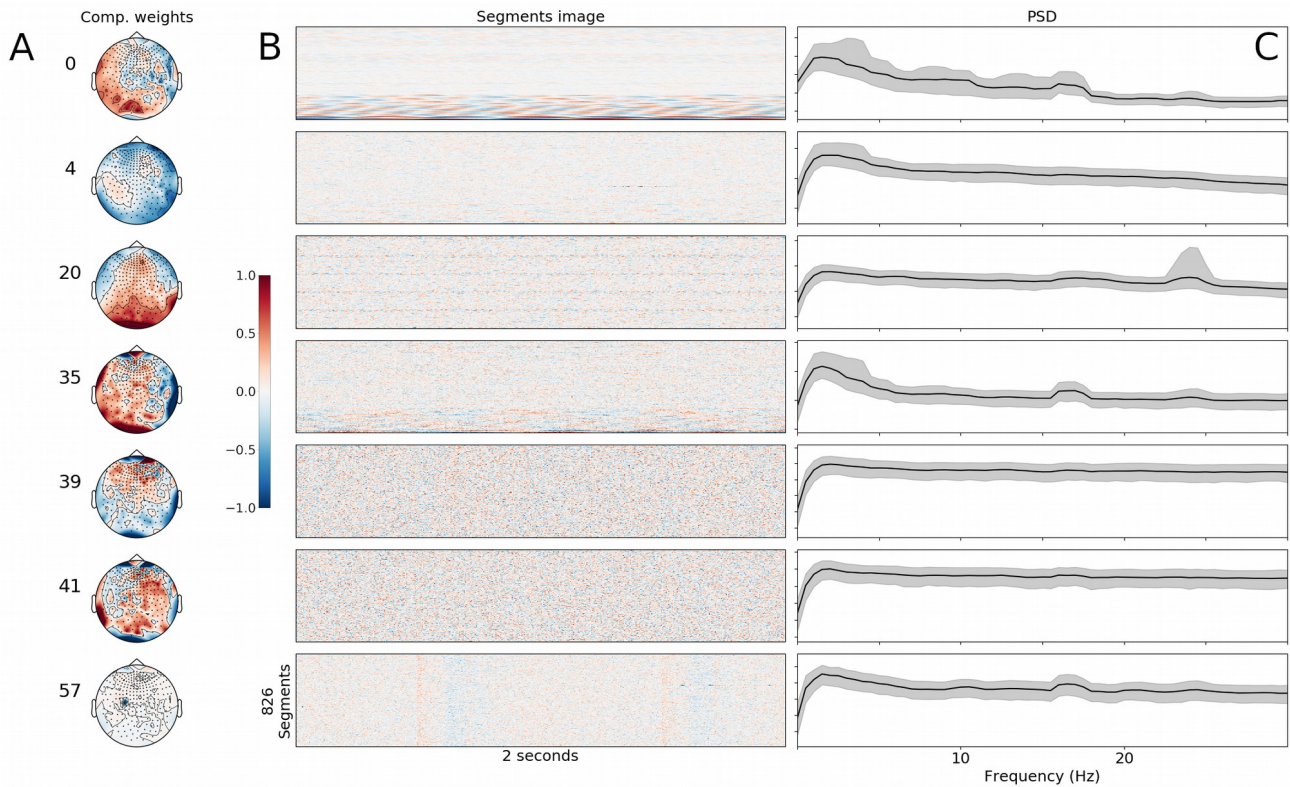

Figure 5: Properties of ICs identified by the "separate" algorithm at Z=3 on the real data example. Component IDs reflect in descending order how much of the variance in the data they explain. A: Topographic representation of the component weights on the 248 normal channels. B: Summary of the components' time courses; each horizontal line in an image represents a two second slice of time in the data. C: Power Spectrum Density (PSD) of the components' derived magnetic fields. Gray areas represent variance across the segments.